\documentclass[aps,prb,twocolumn,showpacs,floatfix]{revtex4}

\usepackage{graphics}
\usepackage{amssymb}

\begin{document}

\title{The Mott transition in the strong coupling perturbation theory}

\author{A. Sherman}
\affiliation{Institute of Physics, University of Tartu, Ravila 14c,
51014 Tartu, Estonia}

\date{\today}

\begin{abstract}
Using the strong coupling diagram technique a self-consistent equation for the electron Green's function is derived for the repulsive Hubbard model. Terms of two lowest orders of the ratio of the bandwidth $\Delta$ to the Hubbard repulsion $U$ are taken into account in the irreducible part of the Larkin equation. The obtained equation is shown to retain causality and gives the correct result in the limit $U\rightarrow 0$. Calculations were performed for the semi-elliptical initial band. It is shown that the approximation describes the Mott transition, which occurs at $U_c=\sqrt{3}\Delta/2$. This value coincides with that obtained in the Hubbard-III approximation. At small deviations from half-filling the density of states shifts along the frequency axis without perceptible changes in its shape. For larger deviations the density of states is modified: it is redistributed in favor of the subband, in which the Fermi level is located, and for $U>U_c$ the Mott gap disappears.
\end{abstract}


\maketitle

\section{Introduction}
The repulsive Hubbard model is one of the main models describing strong electron correlations in crystals. For its investigation a number of methods is used, discussion of which can be found, in particular, in reviews \onlinecite{Georges,Potthoff,Maier,Scalapino}. This paper is devoted to the development of one of such methods -- the strong coupling diagram technique. As follows from its name, the method is aimed at the limit of strong Hubbard repulsion $U$, when it is comparable or larger than the width of the initial electron band $\Delta$. However, due to peculiarities of the method it gives the correct result also in the limit $U\rightarrow 0$ thereby providing an interpolation between the two limits. In calculating Green's functions the method uses the serial expansion in powers of hopping constants. The elements of the arising diagram technique are site cumulants of different orders, connected by hopping lines. As in the diagram technique with the expansion in powers of an interaction, in the present approach the linked-cluster theorem allows one to discard disconnected diagrams and to carry out partial summations in connected diagrams. As a result the one-particle Green's function is expressed in the form of the Larkin equation through the initial electron dispersion and the sum of all irreducible diagrams -- the diagrams, which cannot be divided into two disconnected parts by cutting some hopping line. For more details see Refs.~\onlinecite{Vladimir,Metzner,Pairault,Sherman06,Sherman07}. A somewhat different diagram technique, which is also based on the serial expansion in powers of hopping constants, was developed in Refs.~\onlinecite{Zaitsev,Izyumov,Izyumov90,Ovchinnikov}.

In this paper we consider one-site contributions of the first two orders to the mentioned sum of irreducible diagrams. Due to the restriction to one-site diagrams the considered approximation resembles the dynamic mean field approximation.\cite{Georges} With the insertion of the first- and second-order diagrams into internal hopping lines, the used approximation gives an equation for self-consistent determination of the electron Green's function. It is shown that the solution of this equation retains causality. Calculations were carried out for a semi-elliptical initial band. The approximation is able to describe the Mott transition, which takes place at $U_c=\sqrt{3}\Delta/2$. This value coincides with that found in the Hubbard-III approximation.\cite{Hubbard64} However, in contrast to this latter approximation the obtained density of states (DOS) reduces to the initial semi-elliptical DOS in the limit $U\rightarrow 0$. No quasiparticle peak, inherent in the dynamic mean field approximation, is observed in our calculations at the Mott transition. At small deviations from half-filling the DOS shifts along the frequency axis without perceptible change in its shape. For larger deviations the DOS is redistributed in favor of the subband, in which the Fermi level is located, and for $U>U_c$ the Mott gap disappears. Possible ways of the inclusion of spin and charge fluctuations into the theory are briefly discussed.

\section{Main formulas}
We shall consider the repulsive Hubbard model on a two-dimensional (2D) square lattice. The model is described by the Hamiltonian
\begin{equation}\label{Hamiltonian}
H=\sum_{\bf nn'\sigma}t_{\bf nn'}a^\dagger_{\bf n\sigma}a_{\bf n'\sigma} +\frac{U}{2}\sum_{\bf n\sigma}n_{\bf n\sigma}n_{\bf n,-\sigma},
\end{equation}
where $t_{\bf nn'}$ is the hopping constants, the operator
$a^\dagger_{\bf n\sigma}$ creates an electron on the site {\bf n} of
the 2D lattice with the spin projection $\sigma=\pm 1$ and the electron number operator $n_{\bf n\sigma}=a^\dagger_{\bf n\sigma}a_{\bf n\sigma}$.
In this work we shall calculate the electron Green's function
\begin{equation}\label{Green}
G({\bf n'\tau',l\tau})=\langle{\cal T}\bar{a}_{\bf n'\sigma}(\tau')
a_{\bf n\sigma}(\tau)\rangle,
\end{equation}
where the angular brackets denote the statistical averaging with the
Hamiltonian ${\cal H}=H-\mu\sum_{\bf n\sigma}n_{\bf n\sigma}$, $\mu$ is
the chemical potential, ${\cal T}$ is the time-ordering operator which
arranges operators from right to left in ascending order of times $\tau$,
$a_{\bf n\sigma}(\tau)=\exp({\cal H}\tau)a_{\bf n\sigma}\exp(-{\cal H}\tau)$ and $\bar{a}_{\bf n\sigma}(\tau)=\exp({\cal H}\tau)a^\dagger_{\bf n\sigma}\exp(-{\cal H}\tau)$. Green's function (\ref{Green}) does not depend on the spin projection, and it was omitted in the function notation.

\begin{figure}
\centerline{\resizebox{0.75\columnwidth}{!}{\includegraphics{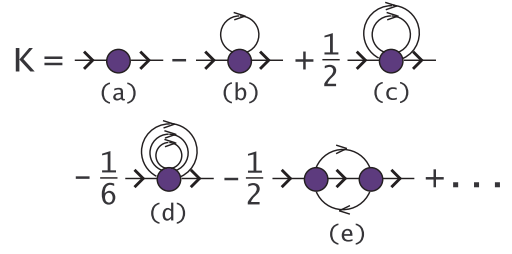}}}
\caption{Irreducible diagrams of the first four orders in the irreducible part $K({\bf k},i\omega_l)$.} \label{Fig1}
\end{figure}

In the strong coupling diagram technique Green's function (\ref{Green}) is presented as the serial expansion in powers of the kinetic term in the Hamiltonian, and the role of the unperturbed Hamiltonian is played by the repulsion term of Eq.~(\ref{Hamiltonian}) together with the term containing the chemical potential,
\begin{equation}\label{unperturbed}
{\cal H}_0=\sum_{\bf n}{\cal H}_{\bf n},\quad {\cal H}_{\bf n}=\sum_\sigma \left(\frac{U}{2}n_{\bf n\sigma}n_{\bf n,-\sigma}-\mu n_{\bf n\sigma}\right).
\end{equation}
Terms of the series are constructed from the hopping constants $t_{\bf nn'}$ and cumulants of the operators $a_{\bf n\sigma}(\tau)$ and $\bar{a}_{\bf n\sigma}(\tau)$ belonging to the same site. The cumulants are calculated with  the site Hamiltonian ${\cal H}_{\bf n}$, Eq.~(\ref{unperturbed}). The sum of all terms of the series can be written in the form of the Larkin equation
\begin{equation}\label{Larkin}
G({\bf k},i\omega_l)=\frac{K({\bf k},i\omega_l)}{1-t_{\bf k}K({\bf k},i\omega_l)},
\end{equation}
where the Fourier transformation over the space and time variables was performed, ${\bf k}$ is the 2D wave vector, $\omega_l=(2l+1)\pi T$ is the Matsubara frequency with the temperature $T$, $t_{\bf k}=\sum_{\bf n}\exp[i{\bf k(n-n')}]t_{\bf nn'}$ and $K({\bf k},i\omega_l)$ is the sum of all irreducible diagrams, which is termed the irreducible part of the Larkin equation. Terms of lowest orders in this sum are shown in Fig.~\ref{Fig1} together with their signs and prefactors. Here circles denote cumulants, which orders equal to numbers of incoming or outgoing directed lines. Two outer arrows designate the operators $\bar{a}_{\bf n'\sigma}(\tau')$ and $a_{\bf n\sigma}(\tau)$ of Green's function (\ref{Green}) in cumulants. The discussed diagram technique allows a partial summation. Therefore, it is presumed that irreducible diagrams of all orders and in all possible combinations are inserted in the internal hopping lines -- the arrowed lines in the diagrams in Fig.~\ref{Fig1}. As a result, in the diagrams, the bare hopping $t_{\bf k}$ is substituted by the renormalized one
\begin{equation}\label{hopping}
\theta({\bf k},i\omega_l)=\frac{t_{\bf k}}{1-t_{\bf k}K({\bf k},i\omega_l)} =t_{\bf k}+t_{\bf k}^2 G({\bf k},i\omega_l).
\end{equation}
The sign of a term in the irreducible part $K({\bf k},i\omega_l)$ is equal to $(-1)^L$, where $L$ is the number of loops formed by hopping lines. Prefactors in the diagrams arise due to the fact that some permutations of the kinetic energy Hamiltonians in a power expansion term lead to the same diagram, since creation and annihilation operators of these permuted Hamiltonians enter into the same cumulant.

The first-order cumulant coincides with the one-particle site Green's function. A cumulant of an order $\nu>1$ is equal to the $\nu$-particle site Green's function and a sum of all possible products of lower-order cumulants, the summarized orders of which is equal to $\nu$. The sign, with which a term appears in this sum, is equal to $-(-1)^P$, where $P$ is the number of permutations performed to obtain the order of operators in the term from that in the $\nu$-th cumulant. In other words, the signs of the terms containing products of first-order cumulants are opposite to signs of the same terms, which would arise from the $\nu$-particle Green's function in the $\nu$-th cumulant if Wick's theorem\cite{Mahan} could be applicable to it. As a result in the limit $U\rightarrow 0$ all cumulants of orders $\nu>1$ vanish, and with them all irreducible diagrams in $K({\bf k},i\omega_l)$, except the diagram (a) in Fig.~\ref{Fig1}, become equal to zero. The first-order cumulant in this latter diagram reads
\begin{eqnarray}\label{C_1}
C_1(i\omega_l)&=&-\int_0^\beta d\tau e^{i\omega_l\tau}\langle a_{\bf n\sigma}(\tau)\bar{a}_{\bf n\sigma}\rangle\nonumber\\
&=&\frac{1}{Z}\Big[\left(e^{-\beta E_1}+e^{-\beta E_0}\right)g_{01}(i\omega_l)\nonumber\\
&&\quad\quad+\left(e^{-\beta E_2}+e^{-\beta E_1}\right)g_{12}(i\omega_l)\Big],
\end{eqnarray}
where $\beta=1/T$, $E_0=0$, $E_1=-\mu$ and $E_2=U-2\mu$ are eigenvalues of the site Hamiltonian $H_{\bf n}$, Eq.~(\ref{unperturbed}), the partition function $Z=e^{-\beta E_0}+2e^{-\beta E_1}+e^{-\beta E_2}$, $g_{ij}(i\omega_l)=(i\omega_l+E_i-E_j)^{-1}$, $i$ and $j=0$, 1 and 2. Substituting $(i\omega_l+\mu)^{-1}$, the value of $C_1(i\omega_l)$ at $U=0$, into Eq.~(\ref{Larkin}) we obtain the correct expression for Green's function of uncorrelated electrons. Thus, the considered approach has an important property -- despite the fact that Eq.~(\ref{Larkin}) was derived from the expansion in powers of $t_{\bf nn'}/U$, it gives the correct result in the limit $U\rightarrow 0$.

In the following consideration we take into account only the one-site diagrams (a) and (b) in Fig.~\ref{Fig1}. Thus, in Eq.~(\ref{Larkin}) the irreducible part does not depend on the wave vector and is given by the equation
\begin{eqnarray}\label{K}
&&K(i\omega_l)=C_1(i\omega_l)\nonumber\\
&&\quad-\frac{T}{N}\sum_{{\bf k}l'\sigma'}C_2(i\omega_l,\sigma;i\omega_l,\sigma;i\omega_{l'},\sigma'; i\omega_{l'},\sigma')\nonumber\\
&&\quad\quad\times\theta({\bf k},i\omega_{l'}),
\end{eqnarray}
where $N$ is the number of lattice sites and $C_2$ is the Fourier transform of the second-order cumulant
\begin{eqnarray*}
&&C_2(\tau_1\sigma,\tau_2\sigma,\tau_3\sigma',\tau_4\sigma')\nonumber\\
&&\quad=\langle{\cal T}\bar{a}_{\bf n\sigma}(\tau_1)a_{\bf n\sigma}(\tau_2)\bar{a}_{\bf n\sigma'}(\tau_3)a_{\bf n\sigma'}(\tau_4)\rangle\\
&&\quad\quad-\langle{\cal T}\bar{a}_{\bf n\sigma}(\tau_1)a_{\bf n\sigma}(\tau_2)\rangle\langle{\cal T}\bar{a}_{\bf n\sigma'}(\tau_3)a_{\bf n\sigma'}(\tau_4)\rangle\\
&&\quad\quad+\langle{\cal T}\bar{a}_{\bf n\sigma}(\tau_1)a_{\bf n\sigma'}(\tau_4)\rangle\langle{\cal T}\bar{a}_{\bf n\sigma'}(\tau_3)a_{\bf n\sigma}(\tau_2)\rangle.
\end{eqnarray*}
The diagrams (c), (d) and other one-site diagrams contain higher orders of the formally small parameter $t_{\bf nn'}/U$. Therefore, they are omitted. The one-site diagrams describe the influence of correlations on the electron spectrum, and in this sense this one-site approximation for $K({\bf k},i\omega_l)$ resembles the dynamic mean field approximation. The diagram (e) is the first one in the series of ladder diagrams, which describe the interaction of electrons with the spin ordering, spin and charge fluctuations. At low temperatures the antiferromagnetic spin order with a large correlation length leads to the doubling of the crystal elementary cell. This manifests itself in some similar features in electron spectral functions for momenta ${\bf k}$ and ${\bf k}+\left(\pi/a,\pi/a\right)$, $a$ being the lattice spacing. In some approximation the full sequence of ladder diagrams can be summed analytically.\cite{Sherman07} These diagrams will be considered in the future work.

In general case the expression for the second-order cumulant, which enters into Eq.~(\ref{K}), is rather cumbersome.\cite{Sherman06} However, it is significantly simplified for the case of principal interest $U\gg T$. If $\mu$ satisfies the conditions
\begin{equation}\label{condition}
T\ll\mu\ll U-T
\end{equation}
the cumulant can be written in the form
\begin{eqnarray}\label{C_2}
&&\sum_{\sigma'}C_2(i\omega_l,\sigma;i\omega_l,\sigma;i\omega_{l'}, \sigma';i\omega_{l'},\sigma')=-\frac{3}{4}\beta F^2(i\omega_l)\delta_{ll'}\nonumber\\
&&\quad\quad+\frac{1}{2} F(i\omega_l)g^2_{01}(i\omega_{l'})+\frac{1}{2} F(i\omega_{l'})g^2_{01}(i\omega_l)\nonumber\\
&&\quad\quad-\frac{1}{2}F(i\omega_l)F(i\omega_{l'})\left[g_{12}(i\omega_l) +g_{12}(i\omega_{l'})\right],
\end{eqnarray}
where $F(i\omega_l)=g_{01}(i\omega_l)-g_{12}(i\omega_l)$. The first-order cumulant (\ref{C_1}) is also somewhat simplified for these conditions,
$$C_1(i\omega_l)=\frac{1}{2}\left[g_{01}(i\omega_l)+
g_{12}(i\omega_l)\right].$$

Combining this expression with Eqs.~(\ref{Green}), (\ref{hopping}), (\ref{K}) and (\ref{C_2}) we obtain a self-consistent equation for calculating $G({\bf k},i\omega_l)$
\begin{eqnarray}
G({\bf k},i\omega_l)&=&\frac{1}{t_{\bf k}}\Bigg[-1+\Bigg(1-t_{\bf k}\bigg\{\frac{1}{2}\left[g_{01}(i\omega_l)+g_{12}(i\omega_l)\right] \nonumber\\
&&\quad\quad+\frac{3}{4}F^2(i\omega_l)\varphi(i\omega_l)
-\frac{s_1}{2}F(i\omega_l)\nonumber\\
&&\quad\quad-\frac{s_2}{2}J(i\omega_l)
\bigg\}\Bigg)^{-1}\Bigg],\label{G_kw}
\end{eqnarray}
where $\varphi(i\omega_l)=N^{-1}\sum_{\bf k}t^2_{\bf k}G({\bf k},i\omega_l)$, $J(i\omega_l)=g^2_{01}(i\omega_l)-F(i\omega_l) g_{12}(i\omega_l)$ and
\begin{equation}\label{s_i}
s_1=T\sum_l J(i\omega_l)\varphi(i\omega_l),\; s_2=T\sum_l F(i\omega_l)\varphi(i\omega_l).
\end{equation}
This equation can be used for calculating Green's function for arbitrary initial dispersion $t_{\bf k}$, for example, by iteration.

Let us perform the analytic continuation to real frequencies $\omega$ and calculate the imaginary part of $G({\bf k}\omega)$. Notice that sums (\ref{s_i}) are real, and the only source of imaginary values is $\varphi(\omega)$. Thus,
\begin{eqnarray}\label{ImG}
&&{\rm Im}\,G({\bf k}\omega)=\frac{3}{4}F^2(\omega){\rm Im}\,\varphi(\omega)\Bigg[\bigg(1+t_{\bf k}\bigg\{\frac{1}{2}F(\omega)s_1 \nonumber\\
&&\quad\quad+\frac{1}{2}J(\omega)s_2-\frac{1}{2}\big[g_{01}(\omega)+ g_{12}(\omega)\big]\nonumber\\
&&\quad\quad-\frac{3}{4}F^2(\omega){\rm Re}\,\varphi(\omega)\bigg\}\bigg)^2
\nonumber\\
&&\quad\quad+\bigg(\frac{3}{4}t_{\bf k}F^2(\omega){\rm Im}\,\varphi(\omega)\bigg)^2
\Bigg]^{-1}.
\end{eqnarray}
As follows from this equation, ${\rm Im}\,G({\bf k}\omega)$ in the left-hand side will be negative if ${\rm Im}\,\varphi(\omega)$ in the right-hand side is negative. Thus, using iteration and substituting in the right-hand side of Eq.~(\ref{ImG}) some function $G_0({\bf k}\omega)$, which is analytic in the upper frequency half-plane, one obtains a function with the same analytic property in the left-hand side, i.e. a retarded Green's function. Consequently, the above equations retain causality.

The problem is essentially simplified in the case of the semi-elliptical initial band with the DOS
\begin{equation}\label{semi-elliptical}
\rho_0(\omega)=\frac{4}{\pi\Delta}\sqrt{1- \left(\frac{2\omega}{\Delta}\right)^2}.
\end{equation}
Thanks to the fact that in the considered approximation the momentum dependence appears in above formulas only through $t_{\bf k}$, summations over the Brillouin zone in these formulas can be performed analytically with this DOS. Let us consider the quantity $\bar{G}(\omega)=N^{-1}\sum_{\bf k}G({\bf k},\omega)$, which is connected with the DOS by the relation
\begin{equation}\label{DOS}
\rho(\omega)=-\pi^{-1}{\rm Im}\,\bar{G}(\omega).
\end{equation}
Passing to dimensionless variables we get from Eq.~(\ref{G_kw})
\begin{equation}\label{Gamma}
\Gamma^5+2\Gamma^3+p_1\Gamma^2+p_2\Gamma+p_1=0,
\end{equation}
where
\begin{eqnarray}
&&\Gamma(f)=\frac{\Delta}{4}\bar{G}(\omega),\quad f=\frac{2}{\Delta}\omega,\nonumber\\
&&v=\frac{U}{\Delta},\quad \lambda=\frac{2}{\Delta}\mu,\quad s'_2=\frac{2}{\Delta}s_2,\nonumber\\
&&p_1=\Big\{2[f+\lambda+v(s_1-1)](f+\lambda)(f+\lambda-2v)
\label{dimensionless}\\
&&\quad\quad -s'_2\left[(f+\lambda-2v)^2+2v(f+\lambda)\right]\Big\}\left(3v^2\right)^{-1}, \nonumber\\
&&p_2=1-\frac{4(f+\lambda)^2(f+\lambda-2v)^2}{3v^2}.\nonumber
\end{eqnarray}
We set $N^{-1}\sum_{\bf k}t_{\bf k}=t_{\bf nn}=0$ and applied this relation in the derivation of Eq.~(\ref{Gamma}).

\section{Results and discussion}
Let us first consider the case of half-filling, $\mu=U/2$, $\lambda=v$. In this case the sums (\ref{s_i}) vanish, since $F(i\omega_l)$ and $J(i\omega_l)$ are even functions of $\omega_l$, while $\varphi(i\omega_l)$ is an odd one. Equation (\ref{Gamma}) is simplified to
\begin{eqnarray}\label{halffilling}
&&\Gamma^5+2\Gamma^3+\frac{2f(f^2-v^2)}{3v^2}\Gamma^2
+\left(1-\frac{4(f^2-v^2)^2}{3v^2}\right)\Gamma\nonumber\\
&&\quad+ \frac{2f(f^2-v^2)}{3v^2}=0.
\end{eqnarray}
A similar equation, however with somewhat different coefficients, was derived in Ref.~\onlinecite{Izyumov}. The difference is connected with neglecting diagrams with spin propagators in this latter work.

Let us consider conditions for the appearance of the Mott gap at the Fermi level $f=0$. For this frequency Eq.~(\ref{halffilling}) is reduced to
\begin{equation}\label{f_eq_0}
\Gamma^5+2\Gamma^3+\left(1-\frac{4}{3}v^2\right)\Gamma=0,
\end{equation}
which has the following five solutions:
\begin{equation}\label{solutions0}
\Gamma=0,\quad \Gamma=\pm\sqrt{-1\pm\frac{2v}{\sqrt{3}}}.
\end{equation}
The solution $\Gamma=-\sqrt{-1+2v/\sqrt{3}}$ is of interest, since for $v<v_c=\sqrt{3}/2$ the quantity $\Gamma(f=0)$ is imaginary, which corresponds to a finite DOS at the Fermi level [see Eqs.~(\ref{DOS}) and (\ref{dimensionless})], while for $v>v_c$ it is real, which means zero DOS. Thus, this solution describes the metal-insulator transition with the value of the critical repulsion
\begin{equation}\label{critical_U}
U_c=\frac{\sqrt{3}}{2}\Delta.
\end{equation}
This value coincides exactly with that obtained in the Hubbard-III approximation.\cite{Hubbard64}

Another analytical solution of Eq.~(\ref{halffilling}) can be obtained for frequencies
\begin{equation}\label{f_divergence}
f=\pm v\quad\rightarrow\quad\omega=\pm\frac{U}{2}.
\end{equation}
For these frequencies Eq.~(\ref{halffilling}) reduces to the equation
\begin{equation}\label{f_eq_v}
\Gamma^5+2\Gamma^3+\Gamma=0
\end{equation}
with the solutions
\begin{equation}\label{solutionsv}
\Gamma=0,\quad\Gamma=\pm i,
\end{equation}
two latter solutions being doubly degenerate. Of these two the solution $\Gamma=-i$ is of interest, since it corresponds to a finite and positive DOS.
Notice that for this solution $K(\omega)$ diverges at $\omega=\pm U/2$. Indeed, from Eq.~(\ref{Larkin}) for the dimensionless irreducible part  $\kappa(f)=\Delta K(f)/2$ we find the relation
\begin{equation}\label{k_vs_G}
\kappa=\frac{2\Gamma}{1+\Gamma^2},
\end{equation}
from which the above statement follows.

\begin{figure}
\centerline{\resizebox{0.85\columnwidth}{!}{\includegraphics{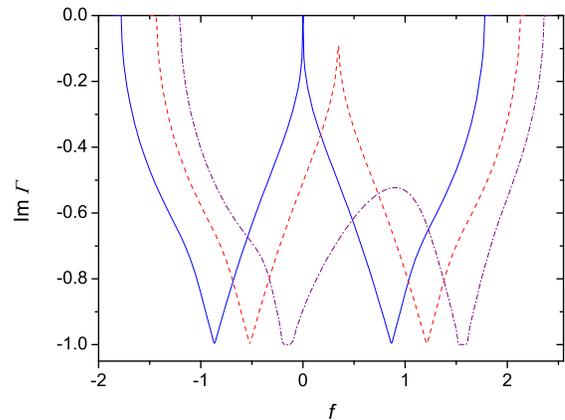}}}
\caption{${\rm Im}\,\Gamma=-\pi\Delta\rho/4$ as a function of $f=2\omega/\Delta$ for $v=U/\Delta=0.866025\approx\ v_c$ and $\mu=0.5U$ (half-filling, $x=1$, the blue solid line), $0.3U$ ($x=0.88$, the red dashed line) and $0.1U$ ($x=0.67$, the purple dash-dotted line). $T=0.001U$.} \label{Fig2}
\end{figure}

\begin{figure}
\centerline{\resizebox{0.85\columnwidth}{!}{\includegraphics{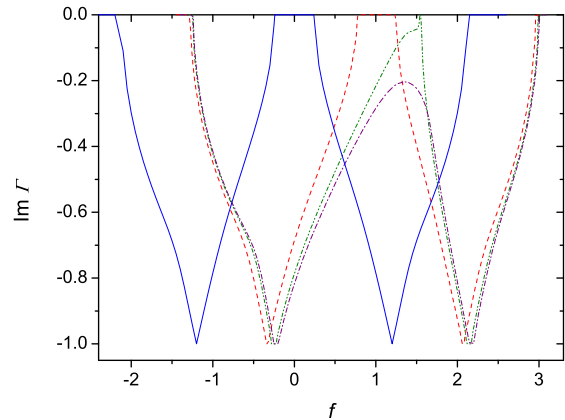}}}
\caption{Same as in Fig.~\protect\ref{Fig2} for $v=U/\Delta=1.2 > v_c$ and $\mu=0.5U$ (the blue solid line), $0.14U$ ($x=0.75$, the red dashed line), $0.11U$ ($x=0.69$, the olive dash-dot-dotted line) and $0.1U$ ($x=0.65$, the purple dash-dotted line). $T=0.001U$.} \label{Fig3}
\end{figure}

\begin{figure}
\centerline{\resizebox{0.85\columnwidth}{!}{\includegraphics{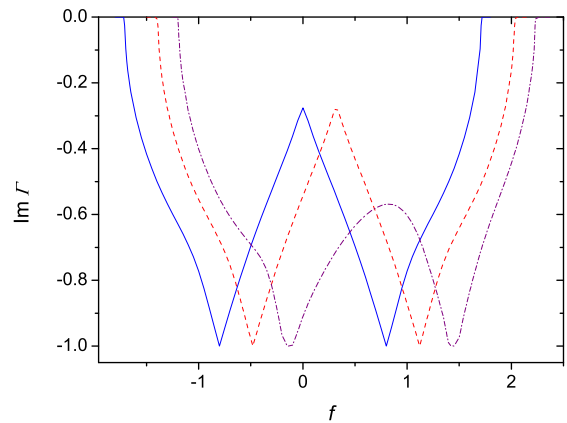}}}
\caption{Same as in Fig.~\protect\ref{Fig2} for $v=U/\Delta=0.8 < v_c$ and $\mu=0.5U$ (the blue solid line), $0.3U$ ($x=0.84$, the red dashed line) and $0.1U$ ($x=0.68$, the purple dash-dotted line). $T=0.001U$.} \label{Fig4}
\end{figure}

The solution of Eq.~(\ref{halffilling}) for arbitrary $f$ is shown in Figs.~\ref{Fig2}--\ref{Fig4} for the cases $v\approx v_c$, $v>v_c$ and $v<v_c$, respectively. As mentioned above, the quantity ${\rm Im}\Gamma$ shown in these figures is connected with the DOS by the relation ${\rm Im}\Gamma=-\pi\Delta\rho/4$ [see Eqs.~(\ref{DOS}) and (\ref{dimensionless})]. To solve Eq.~(\ref{halffilling}) we used the Newton-Raphson method.\cite{Press} As seen from Fig.~\ref{Fig3}, at half-filling for $v>v_c$ near the Fermi level $f=0$ there exists the Mott-Hubbard gap, which magnitude shrinks to zero when $v$ decreases to $v_c$ (Fig.~\ref{Fig2}), and when $v$ becomes less than $v_c$ only a dip at $f=0$ remains (Fig.~\ref{Fig4}). For $v\rightarrow 0$ the DOS approaches its initial value (\ref{semi-elliptical}) with gradual disappearance of the dip. Indeed, in this limit Eq.~(\ref{halffilling}) reduces to the equation
$$\Gamma^2-2f\Gamma+1=0$$
with the solution
$$\Gamma=f-\sqrt{f^2-1},$$
the imaginary part of which coincides with Eq.~(\ref{semi-elliptical}). Thus, as mentioned above, the used approach gives the correct result in the uncorrelated limit $v\rightarrow 0$. Notice that at the metal-insulator transition no quasiparticle peak, which is inherent in the dynamic mean field approximation,\cite{Georges} is observed here.

\begin{figure}
\centerline{\resizebox{0.7\columnwidth}{!}{\includegraphics{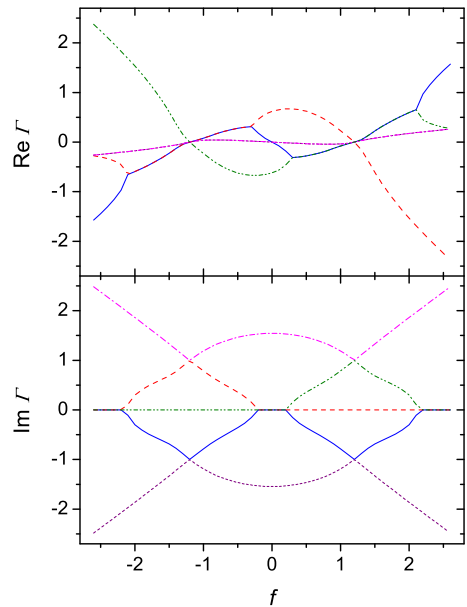}}}
\caption{Frequency dependencies of real and imaginary parts of five solutions of equation (\protect\ref{halffilling}) for $\mu=0.5U$ and $v=1.2$. Different solutions are shown by lines of different colors and types.} \label{Fig5}
\end{figure}

Equation (\ref{halffilling}) is of the fifth order, and together with the solution shown in Figs.~\ref{Fig2}--\ref{Fig4} there are four other solutions. In general these solutions are of no interest -- some of them are real or have positive imaginary parts, which corresponds to zero or negative DOS, respectively (see Fig.~\ref{Fig5} where these solutions are shown by the red dashed, magenta dash-dotted and olive dash-dot-dotted lines). One of these additional solutions have a negative imaginary part. However, this imaginary part spans outside the range of the Mott-Hubbard subbands and remains nonzero near $f=0$ even for $v>v_c$ (the solution shown by the purple short-dashed line in Fig.~\ref{Fig5}). As a consequence this solution does not satisfy the condition $\int_{-\infty}^\infty \rho(\omega)d\omega=1$ even approximately. Notice, that at frequencies (\ref{f_divergence}) this additional solution has the same value $\Gamma=-i$ as the solution of the physical interest [see Eq.~(\ref{solutionsv}) and the following text]. Besides, this latter solution is degenerate with some additional solutions at the edge frequencies of the spectrum. Indeed, Eq.~(\ref{halffilling}) has real coefficients. Therefore, its complex solutions form mutually conjugate pairs. At frequencies corresponding to nonzero DOS the solution of interest has a counterpart with equal in modulus but opposite in sign imaginary part. On the spectrum edges the imaginary parts of both solution vanishes and they become degenerate. Analogous results are obtained in the case $\lambda\neq v$ ($\mu\neq U/2$), which is described by the more general equation (\ref{Gamma}).

In this latter case the solution of Eq.~(\ref{Gamma}) is complicated by the fact that the sums (\ref{s_i}) contain $G({\bf k},i\omega_l)$. In principle, the solution can be found with iteration. However, the problem may be somewhat simplified by using the Hubbard-I approximation\cite{Hubbard63} for $G({\bf k},i\omega_l)$ in estimating $s_1$ and $s_2$. Since Green's function enters into summations in Eq.~(\ref{s_i}), one can expect that a presumably small difference between the function given by the Hubbard-I approximation and the exact one will change the values of these sums only slightly. Green's function in the Hubbard-I approximation is obtained from Eq.~(\ref{Larkin}) if $K({\bf k},i\omega_l)$ is approximated by the first cumulant (\ref{C_1}). With this Green's function, the summation over the Brillouin zone in $\varphi(i\omega_l)$ in Eq.~(\ref{s_i}) can be performed analytically for the DOS (\ref{semi-elliptical}).

In this approximation solutions of Eq.~(\ref{Gamma}) were also obtained with the Newton-Raphson method for different values of $\mu$ and $v$. These solutions are shown in Figs.~\ref{Fig2}--\ref{Fig4}. The captures of these figures contain values of electron concentrations $x$, which was calculated from the obtained ${\rm Im}\Gamma(f)$. Only dependencies for $\mu\leq U/2$ are shown. Due to the symmetry of the problem the curve for $\mu>U/2$ coincides with the one obtained by the specular reflection of ${\rm Im}\Gamma(f)$ for $\mu'=U/2-(\mu-U/2)<U/2$ in the line $f=0$.

If $s_1$ and $s_2$ are negligibly small and can be dropped in Eq.~(\ref{Gamma}), the equation can be solved analytically for frequencies
\begin{equation}\label{f_divergenceG}
f=-\lambda,\; f=2v-\lambda\quad\rightarrow\quad\omega=-\mu,\; \omega=U-\mu.
\end{equation}
In this case Eq.~(\ref{Gamma}) reduces to Eq.~(\ref{f_eq_v}) with the solutions (\ref{solutionsv}). Apparently Eq.~(\ref{f_divergenceG}) is a generalization of Eq.~(\ref{f_divergence}). It appears that the sums $s_1$ and $s_2$ are really small in a wide range of $\mu$ due to mutual compensation of contributions from ${\rm Re}J(i\omega_l){\rm Re}\varphi(i\omega_l)$ and $-{\rm Im}J(i\omega_l){\rm Im}\varphi(i\omega_l)$ into $s_1$ and analogously for $s_2$. In this situation with changing $\mu$ from $U/2$ the DOS is shifted along the frequency axis without perceptible modification of its shape, as it is seen in Fig.~\ref{Fig3} for $\mu=0.14U$ and in Fig.~\ref{Fig4} for $\mu=0.3U$. The exception is the case $v\approx v_c$, when even a small deviation of $\mu$ from $U/2$ leads to the closure of a small Mott gap ($v>v_c$) or to the decrease of the dip depth ($v<v_c$, see Fig.~\ref{Fig2}, the case $\mu=0.3U$).

For even smaller $\mu$ the absolute values of $s_1$ and $s_2$ grow and the deformation of the DOS shape becomes evident even for values of $v$, which differ significantly from $v_c$. For $v>v_c$ the Mott gap disappears, being substituted by a dip (see Fig.~\ref{Fig3} for $\mu=0.11U$ and $0.1U$). For $v<v_c$ the dip becomes shallower (see Fig.~\ref{Fig4} for $\mu=0.1U$). In both cases the DOS is redistributed in favor of the subband in which the Fermi level is located. The minimal value of ${\rm Im}\Gamma$ remains equal to $-1$, as for larger values of $\mu$. However, the minima take more rounded shape.

\section{Conclusion}
In this work, the self-consistent equation for calculating the electron Green's function of the two-dimensional repulsive Hubbard model was obtained using the strong coupling diagram technique. In this derivation, the full sum of irreducible diagrams $K({\bf k},i\omega_l)$ of the Larkin equation was approximated by the diagrams of the two lowest orders of the ratio of the bandwidth $\Delta$ to the Hubbard repulsion $U$. In the second-order term the internal hopping line is renormalized by inserting these irreducible diagrams in it. It was shown that obtained equation retains causality and gives the correct result in the limit $U\rightarrow 0$ thereby providing an interpolation between cases of the weak and strong coupling. Calculations were performed for the initial semi-elliptical band. It was shown that the model describes the Mott metal-insulator transition, which takes place at $U_c=\sqrt{3}\Delta/2$, the value coinciding with that obtained in the Hubbard-III approximation. At the transition, the quasiparticle peak, which is inherent in the dynamic mean field approximation, is not observed in our calculated density of states. The self-consistent equation has five solutions, of which only one has physical meaning in the major part of the frequency range. However, at frequencies corresponding to spectrum edges and to the maximal density of states this solution appear to be degenerate with other solutions of the equation. With deviation from half-filling the frequency dependence of the density of states at first shifts without perceptible change in its shape. Then, for larger deviations, the density of states is redistributed in favor of the subband, in which the Fermi level is located, and, for $U>U_c$, the Mott gap disappears.

Further development of this theory involves the solution of the self-consistent equation for a more realistic initial electron dispersion and the inclusion of interactions of electrons with the magnetic ordering, spin and charge fluctuations. The former task may be performed with the use of iteration, the latter needs in the summation of ladder diagrams describing the spin and charge susceptibilities. These summation can be performed analytically in some approximation.

\begin{acknowledgements}
This work was supported by the European Regional Development Fund (Centre of Excellence "Mesosystems: Theory and applications", TK114) and by the Estonian Scientific Foundation (grant ETF9371).
\end{acknowledgements}

\end{document}